\documentclass[pra,twocolumn,amsmath,amssymb]{revtex4}

\usepackage{graphicx}

\newcommand{\beq}{\begin{equation}}
\newcommand{\eeq}{\end{equation}}
\newcommand{\beqa}{\begin{eqnarray}}
\newcommand{\eeqa}{\end{eqnarray}}
\newcommand{\ket} [1] {\vert #1 \rangle}
\newcommand{\bra} [1] {\langle #1 \vert}
\newcommand{\braket}[2]{\langle #1 | #2 \rangle}
\newcommand{\proj}[1]{\ket{#1}\bra{#1}}
\newcommand{\mean}[1]{\langle #1 \rangle}

\newtheorem{lemma}{Lemma}

\begin{document}

\title{Noise resistance of adiabatic quantum computation
using random matrix theory}
\author{J\'er\'emie Roland}
\author{Nicolas J. Cerf}
\affiliation{Quantum Information and Communication, 
Ecole Polytechnique, CP 165/59, Universit\'e Libre de Bruxelles,
1050 Brussels, Belgium}

\date{\today}

\begin{abstract}
Besides the traditional circuit-based model of quantum computation, 
several quantum algorithms based on a continuous-time Hamiltonian evolution
have recently been introduced, including for instance continuous-time
quantum walk algorithms as well as adiabatic quantum algorithms. 
Unfortunately, very little is known today on the behavior of these Hamiltonian
algorithms in the presence of noise. Here, we perform a fully analytical
study of the resistance to noise of these algorithms using perturbation
theory combined with a theoretical noise model based on random matrices 
drawn from the Gaussian Orthogonal Ensemble, whose elements
vary in time and form a stationary random process.
\end{abstract}

\maketitle



\section{Introduction}
There has been a growing interest recently for the concept of
Hamiltonian-based quantum algorithms, as opposed to the standard
circuit-based paradigm of quantum computing. The Hamiltonian
algorithms rely on the continuous time evolution of a quantum register
according to the Schr\"odinger equation, and include in particular the
quantum search algorithms by adiabatic evolution \cite{farh00} 
or by continuous quantum walks \cite{farh97,chil03}. 
While these algorithms may be translated into circuit-based algorithms
so that they could be implemented on a ``standard''
quantum computer \cite{rola03:hamsearch,ahar04}, another possibility is to consider a ``continuous'' quantum computer specifically designed
to run this type of algorithms \cite{kami04}. 
For a realistic implementation, it seems therefore crucial 
to investigate how well such a quantum computer would behave
in the presence of noise. Until now, this question has only been addressed
for some specific algorithms subject to some very particular noise. 
For instance, Childs {\em et al.} have considered an
adiabatic quantum algorithm for solving combinatorial problems \cite{farh01}
affected by an error modeled by an extra term which is random 
but deterministically evolves in time \cite{chil01}. While this study
was purely numerical, later on Shenvi {\em et al.} \cite{shen03} 
analytically analyzed the effect of a Markovian stochastic variable perturbing the amplitude of the oracle Hamiltonian  in the specific case of
the analog analogue of Grover's search algorithm \cite{farh98}. 
In contrast, the purpose of the present paper is to derive more generic
results for an Hamiltonian-based algorithm perturbed by a noise
that is described by a stationary gaussian random process. This makes
it possible to carry out a fully analytical treatment of the tolerance to noise, although this is at the price of some (fairly general) assumptions 
on the noise model and of the use of perturbation theory.
\par

This paper is organized as follows. In Section \ref{noise}, 
we describe our theoretical model of noise based on the 
Gaussian Orthogonal Ensemble. In Section \ref{time-independent},
we use perturbation theory to analyze the effect of noise on a 
{\em time-independent} Hamiltonian evolution and apply our results 
to the analog analogue of quantum search. In Section \ref{adiabatic},
we consider the tolerance to noise of a {\em time-dependent} quantum algorithm 
by adiabatic evolution, and then focus on the quantum search 
by local adiabatic evolution. Finally, in Section \ref{conclusion},
we conclude by discussing the scaling of the noise-induced error probability 
as a function of the noise bandwidth.

\section{Noise model}\label{noise}

Suppose we have an ideal (noiseless) Hamiltonian algorithm
based on the Hamiltonian $\bar{H}(t)$,
\beq
i\hbar \frac{d}{dt}\ket{\bar{\psi}(t)}=\bar{H}(t)\ket{\bar{\psi}(t)}.
\eeq
At the end of the computation ($t=T$), we obtain 
the state $\ket{\bar{\psi}(T)}$, which, after measurement, defines
the output of the algorithm. Now, suppose that 
a perturbation $\varepsilon h(t)$ adds to the ideal Hamiltonian:
\beq
H(t)=\bar{H}(t)+\varepsilon h(t).
\eeq
Instead of $\ket{\bar{\psi}(T)}$, we will get at the end of the computation a different state $\ket{\psi(T)}$. The problem in the following will be
to evaluate the error probability
\beq\label{perror}
p_\mathrm{err}=1-|\braket{\bar{\psi}(T)}{\psi(T)}|^2
\eeq
induced by the perturbation.

In order to derive analytical results, we will have to make 
some assumptions on the noise term $\varepsilon h(t)$.
First, we limit ourselves to a noise of small amplitude $\varepsilon\ll 1$,
so that the use of perturbation theory is justified. Second,
we assume that, in any basis $\ket{\varphi_k} (k=,0\ldots,N-1)$ of the $N$-dimensional Hilbert space where the computation takes place,
the matrix elements of $h(t)$ are normal random variables:
\beq
h_{kl}(t)=\bra{\varphi_k}h(t)\ket{\varphi_l}\in\mathcal{N}(0,\sigma_{kl}^2).
\eeq
More specifically, we assume that the matrix $h(t)$ 
is drawn from a Gaussian Orthogonal Ensemble (GOE),
so that the variance $\sigma_{kl}^2$ of its elements 
is defined by $\sigma_{kl}^2=(1+\delta_{kl})\sigma^2$,
where $\sigma^2$ is an overall variance (see \cite{stoeckmann} 
for more details on random matrix ensembles).
Moreover, any two distinct elements of a GOE matrix 
are taken as independent random variables,
\beqa
\mean{h_{kl}(t)h_{k'l'}(t)}&=&0\nonumber\\
&\Longleftrightarrow&\label{independent}\\
(k,l)\neq(k',l')\ &\text{and}&\ (k,l)\neq(l',k').\nonumber
\eeqa
Even though the above assumptions are not based on a specific physical
source of noise, they may be justified by considering that the noise
is generally caused by many independent sources of error which, 
combined together, 
finally result in a random Hamiltonian drawn from a GOE as
a consequence of the central-limit theorem.

Furthermore, we assume that the random matrix elements $h_{kl}(t)$ 
evolve in time as some stationary random process
with an autocorrelation function \cite{davenport}
\beq\label{time}
R(\tau)=\mean{h_{kl}(t+\tau)h_{kl}(t)}.
\eeq
For instance, a very typical noise model we can use
is a white noise with a high-frequency cut-off $\omega_0$
(see for instance \cite{agrawal}), which yields
\beq\label{white-noise}
R(\tau)=\sigma_{kl}^2 \frac{\sin \omega_0\tau}{\omega_0\tau}.
\eeq
However, to be slightly more general, we will only assume 
later on that the autocorrelation function is of the type
\beq
R(\tau)=\sigma_{kl}^2 f(\omega_0\tau),
\eeq
where $f(x)$ verifies $f(-x)=f(x)$, $f(x)\leq f(0)=1$, as well 
as some other regularity conditions (see next sections). Thus,
we only need to assume that $R(\tau)$ is a function of $\omega_0\tau$.

Finally, as we will be interested in the scaling of the perturbed
Hamiltonian-based algorithm as a function of the size of the problem, $N$,
we need to properly define the dependence of the noise term in $N$.
For the scaling analysis to be sensible, we must keep a constant 
signal-to-noise ratio as $N$ increases, that is, the eigenvalues
of $h(t)$ should scale similarly to those of $\bar{H}(t)$.
As a result of Wigner's semi-circular law, we know that
the density of eigenvalues of GOE matrices for $N\gg 1$ 
is given by
\beq
\rho(E)\underset{N\to\infty}{\longrightarrow}\left\{
\begin{array}{cl}
\frac{1}{4\sigma^2\pi}\sqrt{4\sigma^2N-E^2}&\text{if}\ |E|\leq\sqrt{4\sigma^2N}\\
0 & \text{otherwise.}
\end{array}
\right.
\eeq
Therefore, to keep a constant signal-to-noise ratio,
we have to impose that $\sigma^2=\bar{E}^2/4N$,
where $\bar{E}$ is of the order of the eigenvalues of $\bar{H}(t)$.

\section{Time-independent Hamiltonian evolution
with noise}\label{time-independent}
\subsection{Perturbation theory}

Let us study first the simplest case of a time-independent 
Hamiltonian evolution.
The solution of the ideal Schr\"odinger equation is
\beq\label{instantunperturb}
\ket{\bar{\psi}(t)}=\sum_k \bar{b}_k e^{-i\frac{E_kt}{\hbar}}\ket{\varphi_k},
\eeq
where $\ket{\varphi_k}$ and $E_k$ are the eigenstates and eigenvalues of the ideal Hamiltonian
and the amplitudes $\bar{b}_k$ follow from the initial conditions.
By use of perturbation theory, we can study the effect 
of a small time-dependent perturbation $\varepsilon h(t)$
on the ideal Hamiltonian $\bar{H}$.
Expanding the solution of the perturbed equation in the basis formed by the solutions of the non-perturbed equation, that is,
\beq
\ket{\psi(t)}=\sum_k b_k(t) e^{-i\frac{E_kt}{\hbar}}\ket{\varphi_k},
\eeq
and introducing this expression into the Schr\"odinger equation, 
we get
\beq
\dot{b}_k=-i\frac{\varepsilon}{\hbar}\sum_l b_l e^{i\omega_{kl}t}h_{kl}(t),
\eeq
where $\omega_{kl}=(E_k-E_l)/\hbar$. Using the same initial state as for the ideal evolution, i.e., $b_k(0)=\bar{b}_k$, we obtain
the system of equations
\beq
b_k(t)=\bar{b}_k-i\frac{\varepsilon}{\hbar}\sum_l \int_0^t b_l(t_1) e^{i\omega_{kl}t_1}h_{kl}(t_1) dt_1.
\eeq
Using standard perturbation theory (see e.g. \cite{schiff}),
this may be solved iteratively, building step by step the expansion
of $b_k(t)$ in increasing orders in $\varepsilon$.
From this solution, one can derive (an expansion of) the error probability $p_\mathrm{err}$ introduced by the perturbation $\varepsilon h(t)$.
As the matrix elements of $h(t)$ are random variables, so will $p_\mathrm{err}$ be, and we will only have access to its statistics. In particular,
we will focus on its mean $\mean{p_\mathrm{err}}$.
Using our assumption that $h(t)$ is a random matrix drawn from a GOE,
we can show that
\beqa
\mean{p_\mathrm{err}}&=&\varepsilon^2 \left\{\sum_{k,l} |\bar{b}_k|^2 (1-|\bar{b}_l|^2) I_{kl}^{-} 
-\sum_{k\neq l} (\bar{b}_k^* \bar{b}_l)^2 I_{kl}^{+} \right\}\nonumber\\
&&+O(\varepsilon^3),\label{avperror}
\eeqa
where we have introduced the integrals
\beq
I_{kl}^\pm=\frac{\sigma_{kl}^2}{\hbar^2} \iint_0^T dt_1dt_2 e^{i\omega_{kl}(t_1\pm t_2)}f\left(\omega_0(t_1-t_2)\right),
\eeq
which correspond to the coupling between 
the states $\ket{\varphi_k}$ and $\ket{\varphi_l}$ 
that is effected by the perturbation.
Our problem now is to evaluate these integrals. 
We see that they only depend on the the noise model via
the autocorrelation function $f(x)$ and the high-frequency cut-off $\omega_0$, while they depend on each particular instance of the problem
via the spectrum of the ideal Hamiltonian or the frequencies
$\omega_{kl}$ (as well as the computation time $T$). Therefore, 
$I_{kl}^\pm$ vary for different instances of a problem.
However, we can derive some general expressions, 
which remain valid for a fairly large class of problems.

First of all, since $I_{kl}^\pm$ are integrals over a domain of size $T^2$ 
and as the amplitude of their integrand is bounded by $1$,
we immediately see that, whatever the values of $\omega_{kl}$ and $\omega_0$,
we have the upper bound
\beq\label{approxallana}
|I_{kl}^\pm|\leq\frac{\sigma_{kl}^2 T^2}{\hbar^2}.
\eeq
Furthermore, we note that the $I_{kl}^+$ couplings only appear between the eigenstates that are initially populated, and may therefore
be viewed as the interferences caused by the noise between these states.
As there is in general a small and fixed number of 
eigenstates $\ket{\varphi_k}$ that are populated ($\bar{b}_k\neq 0$)
in the algorithm \footnote{In particular, we will see that for the analog quantum
search, there are only two populated levels along the evolution,
namely the ground and first excited states.}, Eq.~(\ref{avperror}) implies 
that there will be a fixed number of $I_{kl}^+$ terms contributing to the expression of $\mean{p_\mathrm{err}}$.
In contrast, the number of $I_{kl}^-$ terms, corresponding the
the coupling of the initially populated states to all others,
will in general grow with the dimension $N$ of the Hilbert space.
Therefore, the scaling of the average error probability $\mean{p_\mathrm{err}}$ 
will mostly depend on the integrals $I_{kl}^-$, 
which is why we now focus on these in what follows.
By changing the integration variables to $u=t_1-t_2$ and $v=t_1+t_2$, we get
\beq
I_{kl}^-=2\frac{\sigma_{kl}^2}{\hbar^2} \int_0^T dv \int_0^v du 
\cos(\omega_{kl}u) f(\omega_0 u),
\eeq
which is the integral of a modulated oscillation.

For a white noise (\ref{white-noise}), we get by direct integration
\beqa
I_{kl}^-&=&\frac{\sigma_{kl}^2}{\hbar^2\omega_0}\left[
\frac{1-\cos(\omega_{kl}-\omega_0)T}{\omega_{kl}-\omega_0}+T\ \text{Si}(\omega_{kl}-\omega_0)T\right.\nonumber\\
&&\left.-\frac{1-\cos(\omega_{kl}+\omega_0)T}{\omega_{kl}+\omega_0}+T\ \text{Si}(\omega_{kl}+\omega_0)T\right],\label{exactfunction}
\eeqa
where $\text{Si}(x)$ is the sine integral function.
Depending on the value of $\omega_0$,  we may consider two limiting regimes:
for a high cut-off frequency $\omega_0 \gg \omega_{kl}$, we get
\beq\label{approxrapidana}
I_{kl}^-=\frac{\sigma_{kl}^2}{\hbar^2\omega_0^2} \,  O\left(\left(1+\frac{\omega_{kl}}{\omega_0}\right)\omega_0T\right),
\eeq
while for a low cut-off frequency $\omega_0 \ll \omega_{kl}$, we have
\beq\label{approxslowana}
I_{kl}^-=\frac{\sigma_{kl}^2}{\hbar^2\omega_{kl}^2} \, O\left(1+\frac{\omega_0}{\omega_{kl}}\right).
\eeq
Although Eqs. (\ref{approxrapidana}) and (\ref{approxslowana})
are only valid, strictly speaking, for a white noise,
we obtain similar results for a general function $f(x)$.
In the high-$\omega_0$ regime, since the autocorrelation function $R(\tau)$
usually tends to zero as $\tau$ increases [i.e., $h_{kl}(t+\tau)$ becomes less and less correlated with $h_{kl}(t)$ for increasing $\tau$],
Eq.~(\ref{approxrapidana}) follows from the approximation
$\cos(\omega_{kl}u)f(\omega_0u)=1+O(\omega_{kl}/\omega_0)$.
In the low-$\omega_0$ regime, we must integrate a rapidly oscillating function
over many periods, which is treated in Appendix~\ref{lemma}.
Under very general regularity conditions on $f(x)$
\footnote{Note that in order to satisfy the hypotheses of lemma \ref{lemma2}
in Appendix~\ref{lemma}, $f(x)$ has to be infinitely differentiable.
While this is generally the case for the widely used noise models such as a white noise, $f(x)$ may be discontinuous in $x=0$ for
some specific models, such as $f(x)=\exp(-|x|)$, which could follow from a Poissonian process. This discontinuity in the first derivative
of $f(x)$ is actually linked to the fact that the spectral density of the noise does not rapidly converge to zero for very high frequencies,
and therefore that the noise has some probability to vary arbitrarily fast. We will not consider this case here, but only 
mention that this would yield a drastically different behavior in the low-$\omega_0$ regime.}, we may use the lemma \ref{lemma2} twice, 
and finally recover Eq.~(\ref{approxslowana}).
It is interesting to note that in the low frequency regime, 
the coupling integral does not depend on the computation time $T$. We will see
that, at least for the algorithms considered here, this causes a very different behavior of the scaling for this regime, as compared to the high-$\omega_0$
regime.

\subsection{Analog quantum search}

Let us recall the principle of Farhi and Gutmann's analog quantum search \cite{farh98}. Suppose that we may apply an oracle Hamiltonian
\beq
H_f=\bar{E}(I-\proj{m}).
\eeq
to the system, with $\bar{E}$ denoting an energy scale of the system.
The problem is to prepare the system 
in the (unknown) solution state $\ket{m}$. 
In \cite{farh98}, it is shown that this may be achieved by preparing
the system in the uniform superposition of all states
\beq
\ket{\psi_0}=\frac{1}{\sqrt{N}}\sum_{k=0}^{N-1}\ket{k}
\eeq
and applying the constant Hamiltonian $\bar{H}=H_0+H_f$, where
\beq
H_0=\bar{E}(I-\proj{\psi_0}),
\eeq
during a time
\beq
T=\frac{\pi\hbar}{2\bar{E}}\sqrt{N}.
\eeq
This results in a quadratic speed-up with respect to a classical search
in an unstructured database of size $N$.

\begin{figure}[htb]
\includegraphics[width=8cm]{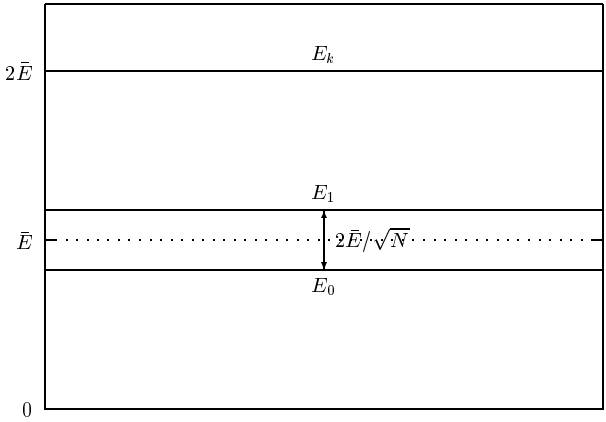}
\caption{Spectrum of the ideal Hamiltonian $\bar{H}=H_0+H_f$.}
\label{spectrumana}
\end{figure}

In order to study the robustness of this quantum algorithm against
a stationary gaussian noise as defined in Sec.~\ref{noise},
let us first consider
the spectrum of the ideal Hamiltonian $\bar{H}=H_0+H_f$ (see Fig.~\ref{spectrumana}). We assume, for simplicity
and without loss of generality, that the problem 
admits the solution $m=0$. The two lowest eigenvalues of $\bar{H}$, 
that is $E_0=(1-x)\bar{E}$ and $E_1=(1+x)\bar{E}$ with $x=1/\sqrt{N}$, are non-degenerate and correspond to the ground and first-excited states,
\beqa
\ket{\varphi_0}&=&\sqrt{\frac{1+x}{2}}\ket{0}+\frac{x}{\sqrt{2(1+x)}}\sum_{k=1}^{N-1}\ket{k}\\
\ket{\varphi_1}&=&\sqrt{\frac{1-x}{2}}\ket{0}-\frac{x}{\sqrt{2(1-x)}}\sum_{k=1}^{N-1}\ket{k},
\eeqa
whereas the $N-2$ times degenerate eigenvalue $E_2=2\bar{E}$ corresponds to the eigenstates
\beq
\ket{\varphi_k}=\frac{1}{\sqrt{2}}(\ket{k}-\ket{1})\quad k=2,\cdots N.
\eeq

Expressing $\ket{\psi(t=0)}=\ket{\psi_0}$ in terms of the eigenstates
$\ket{\varphi_k}$ of the ideal Hamiltonian $\bar{H}$, we get
\beq
\ket{\psi(0)}=\sqrt{\frac{1+x}{2}}\ket{\varphi_0}-\sqrt{\frac{1-x}{2}}\ket{\varphi_1}
\eeq
As a consequence, the instantaneous state of the ideal algorithm $\ket{\bar{\psi}(t)}$ is given by Eq.~(\ref{instantunperturb}) with
\beq
\bar{b}_0=\sqrt{\frac{1+x}{2}},\qquad
\bar{b}_1=-\sqrt{\frac{1-x}{2}},
\eeq
and $\bar{b}_k=0$ for  $k\geq 2$.
Only two states are populated during the ideal algorithm, 
and the average error probability (\ref{avperror}) becomes
\beqa
\mean{p_\mathrm{err}}&=&\varepsilon^2 \left\{(N-2) \left[|\bar{b}_0|^2I_{02}^-+|\bar{b}_1|^2I_{12}^-\right] \right.\nonumber\\
&&+|\bar{b}_0|^2|\bar{b}_1|^2 (I_{00}^-+I_{11}^-)+(|\bar{b}_0|^4+|\bar{b}_1|^4)I_{01}^-\nonumber\\
&&\left.-2\text{Re}\left[(\bar{b}_0^*\bar{b}_1)^2 I_{01}^+\right]\right\}.\label{averroranalog}
\eeqa
where we have used the normalization condition 
$|\bar{b}_0|^2+|\bar{b}_1|^2=1$ and the fact that
$I_{kl}^-=I_{k2}^-$ for $l\ge 2$.
For this algorithm, the bound (\ref{approxallana}) gives
$|I_{kl}^\pm(t)|\leq\pi^2/8$, which is independent of $N$. 
Therefore, only the first term of Eq.~(\ref{averroranalog}), 
which represents the coupling of the
ground and first excited states (the only initially populated states)
to the $N-2$ others, can grow with $N$ and must be taken into account 
in the scaling analysis. Let us focus on this term 
in the two limiting regimes considered above.

For a noise with a high cut-off frequency, Eq.~(\ref{approxrapidana}) yields
\beq
I_{k2}^-=\frac{\bar{E}}{\hbar\omega_0}O\left(\frac{1}{\sqrt{N}}\left(1+\frac{\bar{E}}{\hbar\omega_0}\right)\right)\quad (k=0,1)
\eeq
which is valid if $\hbar\omega_0\gg \bar{E}$.
Clearly, $I_{k2}^-$ 
should be of order $1/N$ for $\mean{p_\mathrm{err}}$ not to grow with $N$, which imposes the condition
\beq  \label{condhighcutoff}
\hbar\omega_0\gg \bar{E}\sqrt{N}.
\eeq
Thus, in this regime, the cut-off frequency of the noise 
must increase as the square root of the size of the problem
in order to keep a probability of error of constant order.
In the case of a noise with a low cut-off frequency,
Eq.~(\ref{approxslowana}) yields
\beq
I_{k2}^-=\frac{1}{N} \, O\left(1+\frac{\hbar\omega_0}{\bar{E}}\right)\quad (k=0,1),
\eeq
so we see that $\mean{p_\mathrm{err}}$ will not grow with $N$
as long as $\hbar\omega_0\ll \bar{E}$. Interestingly, this upper
bound on the cut-off frequency does not depend on the size of the problem.

\begin{figure}
\includegraphics[width=8cm]{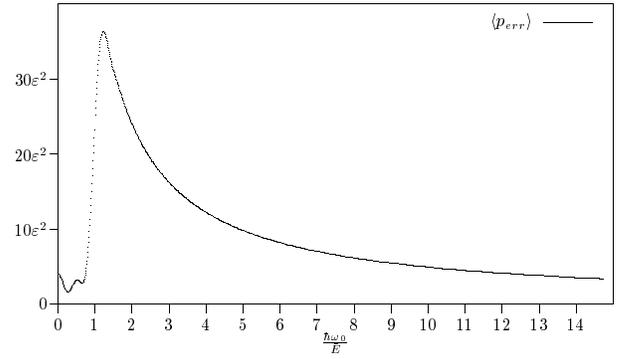}
\caption{Average error probability $\mean{p_\mathrm{err}}$
(to the second order) due to a noise modeled as in Sec.~\ref{noise} with
an autocorrelation function $f(x)=\sin x/x$ 
for the analog search among $N=100$ elements. 
Note that $\mean{p_\mathrm{err}}$ stays very small
as long as $\hbar\omega_0/\bar{E}<1$, at which point 
it shows a sudden increase. For larger values of $\hbar\omega_0$,
it tends progressively back to a low value. The peak at 
$\hbar\omega_0\sim \bar{E}$ scales as $\sqrt{N}$.}
\label{exactgraph}
\end{figure}

We conclude that the influence of noise on the analog quantum search algorithm
is negligible if the noise varies either very slowly or very rapidly with respect to the natural time scale of the problem $\hbar / \bar{E}$.
For the typical case of a white noise with high frequency cut-off, 
the exact integration (\ref{exactfunction}) shows that for intermediate values
of the cut-off frequency $\omega_0 \sim \bar{E}/\hbar$, the average error
probability scales as $\varepsilon^2\sqrt{N}$ 
for a given signal-to-noise ratio $\varepsilon$ (see Fig.~\ref{exactgraph}). 
This means that there exists a forbidden band for the cut-off frequency
$\omega_0$ if we want to keep our algorithm robust with respect
to a constant noise (i.e., a noise that scales with $N$ so to keep a constant
signal-to-noise ratio). Alternatively, we see that the signal-to-noise
ratio $\varepsilon$ should scale as $N^{-1/4}$ in order to keep the error probability $\mean{p_\mathrm{err}}$ constant for a cut-off frequency $\omega_0$
of order $\bar{E}/\hbar$. In other words, 
for an increasing problem size $N$, the noise variance $\sigma^2$ must
decrease faster than $\bar{E}^2/4N$ for the algorithm to remain
immune to noise. Finally,
let us mention that $\varepsilon \sim N^{-1/4}$ coincides with the result of
Shenvi {\em et al.} \cite{shen03}, even though their noise model was 
less general than ours since they only considered an error in the magnitude
of the oracle Hamiltonian, modeled as a Markovian stochastic variable with Gaussian distribution. 

\section{Adiabatic evolution with noise}\label{adiabatic}
\subsection{Adiabatic approximation}

Let us recall the adiabatic approximation, which is
at the basis of the quantum algorithms by adiabatic evolution.
Qualitatively speaking, the idea is as follows: 
if a quantum system is prepared in its ground state and its
Hamiltonian varies ``slowly enough'', it remains in a state 
close to the instantaneous ground state of the Hamiltonian at any time.
To be more precise, let us consider the Schr\"odinger equation for a time-dependent Hamiltonian (see \cite{schiff} for details),
\beq\label{timeschro}
i\hbar \frac{d}{dt}\ket{\bar{\psi}(t)}=\bar{H}(t)\ket{\bar{\psi}(t)}.
\eeq
To solve this equation, we express its solution $\ket{\bar{\psi}(t)}$ in the basis formed by the instantaneous eigenstates $\ket{\varphi_k(t)}$ of the Hamiltonian $\bar{H}(t)$,
\beq
\ket{\bar{\psi}(t)}=\sum_k \bar{b}_k(t) e^{-i\int_0^t\frac{E_k(t_1)}{\hbar}dt_1}\ket{\varphi_k(t)},
\eeq
where $E_k(t)$ are the corresponding instantaneous eigenvalues of $\bar{H}(t)$.
By inserting this expression into Eq.~(\ref{timeschro}), we find
the system of differential equations
\beq
\dot{\bar{b}}_k(t)=\sum_{l\neq k} \bar{b}_l(t) e^{i\int_0^t\omega_{kl}(t_1)dt_1}\frac{\bra{\varphi_k(t)}\frac{d\bar{H}}{dt}\ket{\varphi_l(t)}}{E_k(t)-E_l(t)}.
\eeq
If the quantum system is initially in its ground state $\ket{\bar{\psi}(0)}=\ket{\varphi_0(0)}$, these equations can be integrated,
giving
\beqa  \label{eqnonperturbed}
\lefteqn{ \bar{b}_k(t)=\delta_{0k} }\\
&+&\sum_{l\neq k} \int_0^t\bar{b}_l(t_1) e^{i\int_0^{t_1}\omega_{kl}(t_1')dt_1'}\frac{\bra{\varphi_k(t_1)}\frac{d\bar{H}}{dt_1}\ket{\varphi_l(t_1)}}{E_k(t_1)-E_l(t_1)}dt_1.\nonumber
\eeqa
As in perturbation theory, these equations
may be solved iteratively, which gives after one iteration
\beq  \label{tointegratebyparts}
\bar{b}_k^{(1)}(t)=\int_0^t e^{i\int_0^{t_1}\omega_{k0}(t_1')dt_1'}\frac{\bra{\varphi_k(t_1)}\frac{d\bar{H}}{dt_1}\ket{\varphi_0(t_1)}}{E_k(t_1)-E_0(t_1)}dt_1
\eeq
for $k\neq 0$.
Now if the variation ${d\bar{H}}/{dt}$ of the Hamiltonian is slow enough, or more specifically if
\beq
|A_k(t)| \equiv \hbar\frac{|\bra{\varphi_k(t)}\frac{d\bar{H}}{dt}\ket{\varphi_0(t)}|}{(E_k(t)-E_0(t))^2}\leq\delta_k\ll 1,
\eeq
and under suitable regularity conditions, we may integrate 
Eq.~(\ref{tointegratebyparts}) by parts as shown in Appendix \ref{lemma},
which yields
\beq
\bar{b}_k^{(1)}(t)=
-i\left[A_k(t_1)e^{i\int_0^{t_1}\omega(t')dt'}\right]_0^t+O(\delta_k^2).
\eeq
We see that, at the first order, the amplitudes $|\bar{b}_k(t)|$ 
are bounded by $2\delta_k$.
This first-order (so-called {\em adiabatic}) approximation is acceptable 
if $|\bar{b}_0(t)|^2$ stays close to $1$, that is, if the probability $\bar{p}(t)=\sum_{k\neq 0}|\bar{b}_k(t)|^2$ of hopping to any excited state remains small. In other words, if the adiabatic condition
\beq\label{adiacon}
4\sum_{k\neq 0}\sup_{[0,t]}|A_k(t')|^2\leq\delta^2,
\eeq
is satisfied, then $\bar{p}(t)\le \delta^2 \ll 1$,
with $\delta\ll 1$ being a ``slowness'' parameter.

Now, suppose that a time-dependent perturbation $\varepsilon h(t)$ 
adds to the ideal Hamiltonian $\bar{H}(t)$. 
We again express the solution of the perturbed
Schr\"odinger equation in the basis 
formed by the instantaneous eigenstates of $\bar{H}(t)$,
\beq
\ket{\psi(t)}=\sum_k b_k(t) e^{-i\int_0^t\frac{E_k(t_1)}{\hbar}dt_1}\ket{\varphi_k(t)}
\eeq
which transforms Eq.~(\ref{eqnonperturbed}) into
\beqa
b_k(t)\!\!&=&\!\!\delta_{0k}\\
&+&\!\!\!\!\sum_{l\neq k} \int_0^t b_l(t_1) e^{i\int_0^{t_1}\!\!\omega_{kl}(t_1')dt_1'}\frac{\bra{\varphi_k(t_1)}\frac{d\bar{H}}{dt_1}\ket{\varphi_l(t_1)}}{E_k(t_1)-E_l(t_1)}dt_1\nonumber\\
&-&\!\!\!\!i\frac{\varepsilon}{\hbar}\sum_l \int_0^t b_l(t_1) e^{i\int_0^{t_1}\!\!\omega_{kl}(t_1')dt_1'}\bra{\varphi_k(t_1)}h(t_1)\ket{\varphi_l(t_1)}dt_1.\nonumber
\eeqa
We may again solve this system of equations iteratively, which gives 
after one iteration 
\beqa
\lefteqn{ b_k^{(1)}(t)= \bar{b}_k^{(1)}(t) }  \\
&& -i\frac{\varepsilon}{\hbar} \int_0^t e^{i\int_0^{t_1}\omega_{k0}(t_1')dt_1'}\bra{\varphi_k(t_1)}h(t_1)\ket{\varphi_0(t_1)}dt_1. \nonumber
\eeqa
for $k\neq 0$.
As before, this first-order approximation remains valid provided
that the probability $p(t)=\sum_{k\neq 0}|b_k(t)|^2$
of hopping to any excited state remains small.

Let us now evaluate the average error probability at the end of the evolution
$t=T$ using the same model as before for the perturbation $h(t)$. 
Defining the error probability as $p_\mathrm{err}=p(T)$,
we have
\beqa
\lefteqn{ \mean{p_\mathrm{err}}=\bar{p}_\mathrm{err}+\frac{\varepsilon^2}{\hbar^2}\sum_{k\neq 0}\iint_0^T dt_1dt_2e^{i\int_{t_2}^{t_1}\omega_{k0}(t')dt'} } \nonumber\\
&\times&\mean{\bra{\varphi_k(t_1)}h(t_1)\ket{\varphi_0(t_1)}\bra{\varphi_k(t_2)}h(t_2)\ket{\varphi_0(t_2)}} \nonumber\\
&+&O(\varepsilon^3),
\eeqa
where $\bar{p}_{err}=\bar{p}(T)$ is the error probability of the ideal adiabatic evolution.
Let us note that $\ket{\varphi_k(t_1)}\neq\ket{\varphi_k(t_2)}$ in general, 
so that we do not immediately recover the autocorrelation function of one
particular matrix element of $h(t)$.
However, as the different matrix elements of $h(t)$ are independent in a particular basis, we have
\beqa
\lefteqn{ \mean{\bra{\varphi_k(t_1)}h(t_1)\ket{\varphi_0(t_1)}\bra{\varphi_k(t_2)}h(t_2)\ket{\varphi_0(t_2)}} } \nonumber\\
&=&\left(\braket{\varphi_k(t_2)}{\varphi_k(t_1)}\braket{\varphi_0(t_1)}{\varphi_0(t_2)}\right.\nonumber\\
&&\left.+\braket{\varphi_k(t_2)}{\varphi_0(t_1)}\braket{\varphi_k(t_1)}{\varphi_0(t_2)}\right)\nonumber\\
&&\times \sigma^2_{k0}f(\omega_0(t_1-t_2)).
\eeqa
If $\ket{\varphi_k(t)}$ varies sufficiently smoothly for $0\leq t\leq T$, then the first factor is of order $1-O((t_1-t_2)^2/T^2)$.
Thus, we may approximate it by $1$ as long as $\omega_0T\gg1$, that is,
if the noise varies quickly compared to the adiabatic evolution.
In that case, we get for the average error probability
\beq\label{averroradia}
\mean{p_\mathrm{err}}=\bar{p}_\mathrm{err}+\varepsilon^2\sum_{k\neq 0}I_{k0}^- +O((\delta+\varepsilon)^3),
\eeq
where the integrals
\beq
I_{k0}^-=\frac{\sigma_{k0}^2}{\hbar^2} \iint_0^T dt_1dt_2e^{i\int_{t_2}^{t_1}\omega_{k0}(t')dt'}f(\omega_0(t_1-t_2)),
\eeq
represent the coupling of the ground state to the excited states induced by the perturbation. 
The adiabatic condition generalizes in the case of such a perturbation to
\beq\label{perturbedadiacon}
\sum_{k\neq 0}\left(4\sup_{[0,T]}|A_k(t)|^2+\varepsilon^2I_{k0}^- \right)\leq\delta^2.
\eeq
Similarly to the case of the perturbed time-independent Hamiltonian evolution,
the effect of the perturbation on the adiabatic evolution
mainly depends on the coupling integrals $I_{k0}^-$, which are bounded by
\beq\label{approxalladia}
|I_{k0}^-|\leq\frac{\sigma_{k0}^2 T^2}{\hbar^2}
\eeq
and can be approximated as
\beq\label{approxrapidadia}
I_{k0}^- =\frac{\sigma_{k0}^2}{\hbar^2\omega_0^2} O\left(\left(1+\frac{\omega_{k0}^{\max}}{\omega_0}\right)\omega_0T\right)
\qquad \omega_0\gg \omega_{k0}^{\max}
\eeq
or
\beq\label{approxslowadia}
I_{k0}^- =\frac{\sigma_{k0}^2}{\hbar^2{\omega_{k0}^{\min}}^2} O\left(1+\frac{\omega_0}{\omega_{k0}^{\min}}\right)
\qquad \omega_0\ll \omega_{k0}^{\min}
\eeq
in the limiting regimes of high or low cut-off frequency $\omega_0$,
respectively, 
where $\omega_{k0}^{\min}\leq\omega_{k0}(t)\leq\omega_{k0}^{\max}$
for $t\in [0,T]$.

\subsection{Adiabatic quantum search}

\begin{figure}
\includegraphics[width=8cm]{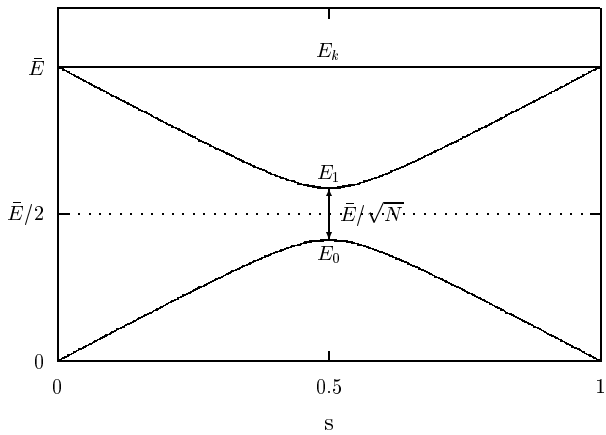}
\caption{Instantaneous eigenvalues of $\tilde{H}(s)$ for $N=32$.}
\label{figure6}
\end{figure}

The principle of the adiabatic quantum search \cite{farh00} 
is to apply the Hamiltonian $H_0=\bar{E}(I-\proj{\psi_0})$ to a system prepared in its ground state $\ket{\psi_0}$ and then to
progressively switch the Hamiltonian $H_0$ to the Hamiltonian $H_f=\bar{E}(I-\proj{m})$, where $m$ is the solution of the search problem. 
If this switch is done slowly enough,
the system will stay in the instantaneous ground state of the Hamiltonian and thus end up in the ground state of $H_f$, i.e., the solution state
$\ket{m}$. The instantaneous Hamiltonian is chosen as
\beq
\tilde{H}(s)=(1-s)H_0+s H_f,
\eeq
where $s=s(t)$ is an evolution function which must be optimized 
so as to reduce the computation time while respecting the adiabatic condition~(\ref{adiacon}) (see \cite{vdam01,rola02} for details).
In this case, this condition may be rewritten as
\beq\label{adiacon2}
\frac{ds}{dt}\leq\frac{\delta}{2\hbar}
\frac{(E_1(t)-E_0(t))^2}{|\bra{\varphi_1(t)}H_f-H_0\ket{\varphi_0(t)}|}.
\eeq
Without loss of generality, we may once again suppose that $m=0$.
The instantaneous eigenstates of $\tilde{H}(s)$ are
\beqa
\ket{\varphi_0(s)}&=&\frac{\sqrt{N}(E_1(s)-s)\ket{\psi_0}+s\ket{0}}{\sqrt{E_1(s)^2+(N-1)(E_1(s)-s)^2}}\\
\ket{\varphi_1(s)}&=&\frac{\sqrt{N}(E_0(s)-s)\ket{\psi_0}+s\ket{0}}{\sqrt{E_0(s)^2+(N-1)(E_0(s)-s)^2}}\\
\ket{\varphi_k(s)}&=&\frac{1}{\sqrt{2}}(\ket{k}-\ket{1})\qquad k\ge 2 ,
\eeqa
where
\beqa
E_0(s)&=&\frac{\bar{E}}{2}\left[1-\sqrt{1-4\frac{N-1}{N}s(1-s)}\right]\\
E_1(s)&=&\frac{\bar{E}}{2}\left[1+\sqrt{1-4\frac{N-1}{N}s(1-s)}\right]\\
E_k(s)&=&\bar{E}\qquad k\ge 2
\eeqa
are the instantaneous eigenvalues of $\tilde{H}(s)$ (see Fig.~\ref{figure6}).
Since $||H_f-H_0||\leq \bar{E}$, taking an evolution function $s(t)$
that satisfies
\beq
\frac{ds}{dt}=\frac{\delta}{2\hbar\bar{E}} (E_1(s)-E_0(s))^2
\eeq
complies with the adiabatic condition~(\ref{adiacon2}), 
and then leads to a computation time
\beq
T=\frac{\pi}{\delta}\frac{\hbar}{\bar{E}}\sqrt{N}.
\eeq

Consider that some noise, modeled again as a stationary gaussian
random process, perturbs the evolution. 
Equation~(\ref{averroradia}) then reads
\beq
\mean{p_\mathrm{err}}\leq\bar{p}_\mathrm{err}
+\varepsilon^2\left[I_{10}^-+(N-2)I_{k0}^-\right]
+O((\delta+\varepsilon)^3),
\eeq
where $I_{k0}^-\leq \pi^2/(64\delta^2)$ as a result of
Eq.~(\ref{approxalladia}).
Let us emphasize that, while it was only the excitation 
to the first excited state that was critical for the ideal adiabatic algorithm,
in this case it is the coupling of the ground state to all excited states that could make the algorithm fail,
since their number grows as the size $N$ of the problem.
Moreover, this bound already suggests 
that the coupling integrals $I_{k0}^-$ 
-- and therefore the error probability -- could increase when the
evolution slows down ($\delta$ decreases) which means there must be a compromise between a slow evolution, very close to perfect adiabaticity, and a fast
evolution, more robust to noise.

As before, let us consider the two limiting regimes of a high or a low
cut-off frequency $\omega_0$.
In the case of a high cut-off frequency ($\hbar\omega_0\gg\bar{E}$),
Eq.~(\ref{approxrapidadia}) yields
\beq
I_{k0}^-=\frac{\bar{E}}{\hbar\omega_0} \, O\left(\frac{1}{\delta\sqrt{N}}\left(1+\frac{\bar{E}}{\hbar\omega_0}\right)\right) \quad \forall\ k\neq 0,
\eeq
exactly as for the analog quantum search except for the factor $1/\delta$.
The latter factor shows that in order to keep the algorithm robust to noise,
the cut-off frequency has to increase not only 
as the the size of the database $N$ grows
(just as for the analog quantum search), but also as the evolution slows down
($\delta$ decreases). More precisely, we see that the perturbed adiabatic
condition (\ref{perturbedadiacon}) is satisfied only if
\beq   \label{condhighcutoff2}
\hbar\omega_0 \gg \frac{\varepsilon^2}{\delta^3} \bar{E} \sqrt{N}
\eeq
(compare with Eq.~(\ref{condhighcutoff})).
When the cut-off frequency becomes very low ($\hbar\omega_0\ll\bar{E}$), 
Eq.~(\ref{approxslowadia}) implies that the coupling integrals behave as
\beq
I_{k0}^-=\frac{1}{N} \, O\left(1+\frac{\hbar\omega_0}{\bar{E}}\right)
\eeq
for all excited states except the first one (i.e., for $k\ge 2$). 
For the first excited state ($k=1$), we have $\hbar \omega_{10}^{\min}
\sim \bar{E} / \sqrt{N}$ so Eq.~(\ref{approxslowadia}) does not yield
a useful result. Instead, we simply use
the general bound $I_{k0}^-\leq \pi^2/(64\delta^2)$. Therefore, 
the adiabatic condition is satisfied here as long as
$\hbar\omega_0\ll\bar{E}$ (just as for the analog quantum search),
but also if $\varepsilon\ll \delta$.

In summary, we recover essentially the same effects for the adiabatic
quantum search as for the analog quantum search, 
that is, the influence of noise becomes negligible only in the case of
a very high or a very low cut-off frequency $\omega_0$, 
apart from the influence of the slowness parameter $\delta$.
Regarding this latter parameter,
we see that while decreasing $\delta$ gets the ideal evolution closer to adiabaticity and therefore reduces the error probability without noise, in the presence of noise it imposes that $\varepsilon$ decreases 
-- i.e., that the signal-to-noise ratio increases -- in both regimes of a high
or low cut-off frequency (or that the high cut-off frequency increases 
as $1/\delta^3$).

\section{Conclusion}\label{conclusion}

We have studied the resistance of Hamiltonian quantum algorithms (including
adiabatic algorithms) to a noise that is modeled as a random matrix
whose elements are stationary gaussian random processes within a fixed
bandwidth. This statistical noise model is generic, and should therefore 
make our analysis valid over a large class of physical systems, regardless
of the exact origin of the added noise. Another main advantage of this
noise model is that it makes it possible to perform a fully analytical 
scaling analysis. Our general result is that the Hamiltonian
algorithms are resistant to noise (i.e., the error probability 
$\mean{p_\mathrm{err}}$ does not increase
with increasing problem sizes $N$) as long as the cut-off frequency
of the noise is either very high or very low
with respect to the inverse of the characteristic time-scale of the system
$\bar{E}/\hbar$. Asides from the influence of the slowness parameter $\delta$
in the case of the adiabatic algorithms, this resistance is essentially 
similar for adiabatic and time-independent Hamiltonian algorithms.
Our results are in good agreement with the numerical study of
Childs {\em et al.} in \cite{chil01}. They even corroborate the results
of Shenvi {\em et al.} \cite{shen03}, although their noise model
was rather different, which supports the idea that using random matrix theory
provides a rather general description of noise.

Roughly speaking, the two limiting regimes of high or low cut-off frequencies
can be understood in the following way.
If the frequency components of the noise are
much below the inverse of the characteristic time-scale of the system,
it is intuitively clear that the noise cannot effect transitions to undesired 
states. On the contrary, if the noise spectrum spreads over a band which
is much broader than the inverse of the characteristic time-scale of the system,
then the noise spectral density is low around the frequencies that effect
undesired transitions. In the intermediate region, we found that
the error probability unfortunately scales as $\sqrt{N}$ \footnote{This scaling has been analytically
obtained for a time-independent Hamiltonian, and is strongly believed to generalize to the case
of an adiabatic evolution.}, which implies that some error correction is needed
to make Hamiltonian algorithms scalable. This last point is particularly
important as it is plausible that the source of noise occurring
in a physical system typically varies on a time scale comparable 
to the natural time scale of the system, so that the less favorable
regime ($\hbar\omega_0 \sim \bar{E}$) may be the most common situation.

Coming back to the two limiting regimes, let us notice
that the high cut-off frequency increases towards higher frequencies 
when $N$ raises [see Eqs. (\ref{condhighcutoff}) and (\ref{condhighcutoff2})].
Consequently, the Hamiltonian algorithms will in practice not be scalable 
in the high cut-off frequency regime too, since the noise should spread 
over an arbitrarily large spectrum to keep the error probability low. 
In this case, some kind of error correction
should also be implemented if the size of the problem becomes too large.
The case of a low cut-off frequency, however, is more favorable.
Indeed, the situation is quite different here as the error probability 
stays small as soon as the cut-off frequency is lower than 
some fixed value, even when the size of the problem increases.
This fault tolerance may be explained by the fact that the spectral density of noise does not contain frequencies close to resonances, and thus will not efficiently couple different eigenstates of the ideal Hamiltonian.

It should be emphasized that it is not the possible excitation to one 
particular state that makes the algorithm fail, but the fact the dimension of the Hilbert space increases with the problem size, and hence the number of states that could be accidentally populated as well.
This means, in the case of adiabatic computation, that even if the gap between the ground and first excited states decreases, the algorithm may remain robust to a noise with a low cut-off frequency (even if this frequency remains constant) as long as the gap between the ground states 
and the other excited states remain lower bounded. Therefore, the algorithm would remain scalable in the
case of a low cut-off frequency as long as the natural frequencies 
of the ideal Hamiltonian are much larger than the frequencies
contained in the noise. Of course, throughout this analysis, we always
made the assumption that the signal-to-noise ratio remains essentially 
constant when the  size of the Hilbert space where the computation 
takes place becomes large, 
which may practically not be the case. Thus,
even in this low cut-off frequency regime, it may be necessary 
to devise error correction techniques for Hamiltonian --
and in particular adiabatic -- quantum algorithms.

{\em Note:} A few days ago, a related paper appeared on the quant-ph preprint
server, which also considers the use of random matrix theory in adiabatic
quantum computing but with a distinct goal, namely to analyze 
the spectral statistics of the Hamiltonian over a large class of problems 
along an adiabatic (but noiseless) evolution \cite{mitc04}.

\begin{acknowledgments}
We acknowledge financial support from the Communaut\'e Fran\c{c}aise de
Belgique under grant ARC 00/05-251, from the IUAP programme of the Belgian
government under grant V-18, and from the EU under projects RESQ
(IST-2001-37559) and CHIC (IST-2001-33578)
J.R. acknowledges support from the Belgian foundation FRIA.
\end{acknowledgments}

\appendix{}
\section{}\label{lemma}
In this appendix, we give a useful tool to evaluate integrals of an oscillating function such as
\beq\label{integral}
\int_a^bdxF(x)e^{i\omega x}.
\eeq
The basic idea relies on Riemann-Lebesgue's lemma:
\begin{lemma}[Riemann-Lebesgue]
If $F(x)$ is an integrable function on $[a,b]$, then
\beq\nonumber
\lim_{\omega\to\infty}\int_a^b dx F(x)e^{i\omega x}=0.
\eeq
\end{lemma}

This lemma suggests that the integral (\ref{integral}) will be {\em relatively small} if $\omega$ is {\em sufficiently large}.
The purpose of this appendix is to quantify this idea.

First of all, as long as $F(x)$ is differentiable on $[a,b]$, we may integrate (\ref{integral}) by parts:
\beq\nonumber
\int_a^bdxF(x)e^{i\omega x}
=-\frac{i}{\omega}\left[F(x)e^{i\omega x}\right]_a^b+\frac{i}{\omega}\int_a^bdx\frac{dF}{dx}(x)e^{i\omega x},
\eeq
where $\left[f(x)\right]_a^b=f(b)-f(a)$, and, using this last equation iteratively, we show that
for an $N$-times differentiable function $F(x)$ on $[a,b]$,
\beqa
\int_a^b dx F(x)e^{i\omega x}&=&
-\sum_{n=0}^{N-1}\left(\frac{i}{\omega}\right)^{n+1}\left[\frac{d^nF}{dx^n}(x)e^{i\omega x}\right]_a^b\nonumber\\
&&+\left(\frac{i}{\omega}\right)^N\int_a^bdx\frac{d^NF}{dx^N}(x)e^{i\omega x}.
\eeqa
The order of the error introduced by neglecting the last term may be evaluated as follows:
\beq
\left|\left(\frac{i}{\omega}\right)^N\int_a^bdx\frac{d^NF}{dx^N}(x)e^{i\omega x}\right|
\leq\frac{1}{\omega^N}\int_a^bdx\left|\frac{d^NF}{dx^N}(x)\right|.
\eeq
We see that the accuracy of this approximation increases with the oscillation frequency $\omega$.
Moreover, if $(1/\omega^N){d^NF}/{dx^N}\to 0$ for $N\to\infty$, this error approaches zero as $N$ increases and we prove the following lemma:
\begin{lemma}\label{lemma2}
Let the function $F(x)$ be infinitely differentiable on $[a,b]$. If
\beq\nonumber
\frac{1}{\omega^N}\frac{d^NF}{dx^N}(x)\underset{N\to\infty}{\longrightarrow} 0\quad\forall\ x
\eeq
for some real $\omega$, then
\beq\nonumber
\int_a^b dx F(x)e^{i\omega x}=
-\sum_{n=0}^{\infty}\left(\frac{i}{\omega}\right)^{n+1}\left[\frac{d^nF}{dx^n}(x)e^{i\omega x}\right]_a^b.
\eeq
\end{lemma}

While this result is helpful to study a time-independent Hamiltonian evolution, in the case of an adiabatic evolution the
typical frequencies become time-dependent.
However, using the same method, we easily generalize this lemma to the case of a varying frequency $\omega(x)$.
\begin{lemma}\label{lemma3}
Let the function $F(x)$ be infinitely differentiable on $[a,b]$. If
\beq\nonumber
\frac{1}{\omega(x)^N}\frac{d^NF}{dx^N}(x)\underset{N\to\infty}{\longrightarrow} 0\quad\forall\ x
\eeq
for some real differentiable function $\omega(x)$ on $[a,b]$, then
\beqa
&&\int_a^b dx F(x)e^{i\int_0^x\omega(x') dx'}\nonumber\\
&=&\sum_{n=0}^{\infty}\left\{-\left[\left(\frac{i}{\omega(x)}\right)^{n+1}\frac{d^nF}{dx^n}(x)e^{i\int_0^x\omega(x') dx'}\right]_a^b\right.\nonumber\\
&&\left.+\int_a^bdx\frac{d}{dx}\left(\frac{i}{\omega(x)}\right)^{n+1}\frac{d^nF}{dx^n}(x)e^{i\int_0^x\omega(x') dx'}\right\}.\nonumber
\eeqa
\end{lemma}

\bibliography{qit,noise}

\end{document}